\documentclass[a4paper,fleqn,usenatbib,letters]{mnras}

\usepackage[T1]{fontenc}
\usepackage{ae,aecompl}

\usepackage[dvipdfmx]{graphicx}	
\usepackage{amsmath}	
\usepackage{amssymb}	
\usepackage[hang,small,bf]{caption}
\usepackage[subrefformat=parens]{subcaption}
\usepackage{color}

\title[Radiation feedback in dusty clouds]{Radiation feedback in dusty clouds}

\author[S. Ishiki \& T. Okamoto]{
Shohei Ishiki,$^{1}$\thanks{E-mail: ishiki@astro1.sci.hokudai.ac.jp}
Takashi Okamoto$^{1}$
\\
$^{1}$Department of Cosmoscience, Hokkaido University, N10 W8, Kitaku, Sapporo, 060-0810, Japan\\
}

\date{Accepted XXX. Received YYY; in original form ZZZ}

\pubyear{2015}

\begin{document}
\label{firsPHage}
\pagerange{\pageref{firsPHage}--\pageref{lasPHage}}
\maketitle

\begin{abstract}
We have investigated the impact of photoionization and radiation pressure on a dusty star-forming cloud 
using one-dimensional radiation hydrodynamic simulations, which include 
absorption and re-emission of photons by dust. 
  We find that, in a cloud of mass $10^{5}~M_{\sun}$ and radius 17~pc, the effect of radiation pressure is negligible 
  when star formation efficiency is 2~\%. 
  The importance of radiation pressure increases with 
  increasing star formation efficiency or an increasing dust-to-gas mass ratio. 
  The net effect of radiation feedback, however, becomes smaller with the increasing 
  dust-to-gas mass ratio, since the absorption of ultra-violet photons by dust grains 
  suppresses photoionization and hence photoheating. 

\end{abstract}

\begin{keywords}
radiative transfer -- methods: numerical -- ISM: clouds -- \ion{H}{ii} regions 
\end{keywords}



\section{Introduction}

Radiative transfer is known to be very important in many astrophysical phenomena. 
Feedback from young, massive stars plays a crucial role in determining star formation 
activity and galaxy evolution. 
The energy and momentum input by stellar radiation from young, massive stars 
is most influential in a star-forming cloud before the explosion of the first 
supernova. 

The radiation from young, massive stars can affect the surrounding medium 
through two channels. 
First, ultraviolet (UV) photons ionize the surrounding neutral gas and increase 
its temperature by photoheating. 
The \ion{H}{ii} region expands owing to the increased thermal pressure. 
This process disperses star-forming clouds \citep{walch12} and changes star formation 
efficiency \citep{2004RvMP...76..125M, 2007ARA&A..45..565M}. 
Secondly, neutral gas and dust absorb photons and acquire their 
momentum. The momentum pushes the material outward. 
This process may drive galaxy scale outflows 
\citep{2005ApJ...618..569M, 2006MNRAS.373.1265O}. 
The ionization feedback \citep{Dale2007,Peters2010,Dale2012,hosokawa2015} 
and radiation pressure feedback \citep{Krumholz2007, Kuiper2010b, Kuiper2011, Kuiper2012, Kuiper2013, Harries2014} 
are also important for individual star formation. 

Recently, \citet{sales14} have studied these processes and have concluded that radiation pressure 
has negligible effect compared with photoheating. 
They, however, do not include dust in their simulations.
On the other hand, \citet{wise12} perform radiation hydrodynamic simulations in full cosmological 
context and show that momentum input partially affect star formation by increasing the turbulent 
support in early low-mass galaxies, while they ignore the dust since dust is 
unimportant in these low metallicity systems.  
 
The presence of dust increases the importance of radiation pressure because, 
unlike hydrogen and helium, dust can always absorb UV photons. 
Moreover, absorbed photons are re-radiated as 
infrared (IR) photons and again dust absorbs IR photons. 
Iterative process of the absorption and the re-emission increases the efficiency of converting 
photon energy to dust momentum. 
The creation of \ion{H}{ii} regions by UV radiation and momentum-driven gas outflows by 
absorption of re-emitted IR photons now become important ingredients in galaxy formation 
simulations, which are modelled as phenomenological subgrid physics 
\citep[e.g.][]{hopkins11, brook12, stinson13, agertz13, okamoto14}. 

It is therefore important to investigate the radiative feedback 
by radiation hydrodynamic simulations that include re-emission from dust grains. 
To this end, we perform one-dimensional radiation hydrodynamic simulations in the presence of dust. 
In this Letter, we solve radiation transfer including angular dependence in order to deal with 
re-emission from dust and gas.  


\section{Methods}
\label{sec:Methods}

We place a radiation source at the centre of a spherically symmetric gas distribution. 
To compare the effect of thermal and radiation pressure, we perform simulations 
with and without radiation pressure and investigate relative importance of these 
processes. 

\subsection{Radiation transfer}
We here describe the algorithm that we use to 
solve the steady radiative transfer equation for a given frequency, $\nu$: 
\begin{equation}
	\frac{dI_{\nu}}{d \tau_{\nu}} = - I_{\nu} + S_{\nu}, 
	\label{eq:radiation_transfer}
\end{equation}
where $I_{\nu}$, $\tau_\nu$, and $S_\nu$ are the specific intensity, the optical depth, and the source function, 
respectively. 
Optical depth of a ray segment, $\Delta \tau_{\nu}$, is determined as 
\begin{equation}
\Delta \tau_{\nu} = \kappa_{\nu} \Delta x=\sum_i n_i \sigma_i \Delta x ,
\end{equation} 
where $\kappa_{\nu}$, $n_i$, and $\sigma_i$ are the absorption coefficient, the number density, 
and the cross section of $i$th species of interest, respectively, and 
$\Delta x$ is the length of the ray segment that intersects the cell. 
The species we include in our simulations are \ion{H}{i}, \ion{H}{ii}, \ion{He}{i}, \ion{He}{ii}, 
\ion{He}{iii}, electrons, and dust.  
We employ the cross-sections of \ion{H}{i}, \ion{He}{i}, and \ion{He}{ii} given in \citet{Osterbrock2006} 
and that of dust in \citet{1984ApJ...285...89D} 
and \citet{1993ApJ...402..441L}\footnote{http://www.astro.princeton.edu/~draine/dust/dust.diel.html}. 

The recombination radiation from ionized hydrogen and helium is calculated as 
\begin{equation}
	\frac{S_{\nu,i}}{\kappa_{\nu}} = \frac{ \alpha_{i} (T)  n_\mathrm{e} n_i h \nu}
	{4 {\mathrm \pi } \sqrt{ {\mathrm \pi } } \Delta \nu_{\mathrm{D},i}}\mathrm{e}^{- (\nu -\nu_0)^2/(\Delta \nu_\mathrm{D})^2}, 
	\label{eq:reemission_gas}
\end{equation}
where 
$h$ is the plank constant, $\alpha_{i}$ is the recombination coefficient 
for a transition from ionized state to ground state, $\nu_{0,i}$ is the threshold frequency of the $i$th species, 
$n_\mathrm{e}$ is the electron number density, $T$ is gas temperature, and $\Delta \nu_{\mathrm{D},i}$ is the Doppler width 
defined as: 
\begin{equation}
\Delta \nu_{\mathrm{D},i} = \frac{\nu_0}{c} \sqrt{\frac{2kT}{m_i}},
\end{equation}
where $k$ is the Boltzmann constant and $c$ is the speed of light.
For spherically symmetric systems, intensities are functions of radius and angle if the problem 
involves re-emission of photons as in our case.  
We therefore employ a scheme called the impact parameter method \citep{HummerRybicki1971}.  

\subsection{Chemical reactions and radiative heating and cooling}

In our simulations, we solve a network of chemistry consists of \ion{H}{i}, \ion{H}{ii}, \ion{He}{i}, 
\ion{He}{ii}, \ion{He}{iii}, and electrons, which can be described by a following set of equations:
\begin{equation}
\frac{dn_i}{dt} = C_i - D_i n_i, 
\label{eq:chemistry}
\end{equation}
where $n_i$ is number density of the $i$th species, $C_i$ is the collective source term responsible for the 
creation of the $i$th species, and the second term involving $D_i$ represents the destruction mechanisms 
for the $i$th species. 
Since equation~(\ref{eq:chemistry}) is a stiff set of differential equations, we need an implicit scheme 
for solving them.
We thus employ a backward difference formula \citep{1997NewA....2..209A, OYU12}: 
\begin{equation}
n^{t + \Delta t}_{i} = \frac{C^{t + \Delta t}_i + n^{t}_{i}}{1 + D^{t + \Delta t}_i \Delta t }, 
\end{equation}
where $C^{t + \Delta t}_i$ and $D^{t + \Delta t}_i$ are evaluated at the advanced timestep. 
Unfortunately, not all source terms can be evaluated at the advanced timestep due to 
the intrinsic non-linearity of equation~(\ref{eq:chemistry}).  
We thus sequentially update the number densities of all species in the order of increasing 
ionization states.

The chemical reactions included in our simulations are the recombination 
\citep{1994MNRAS.268..109H, 1998MNRAS.297.1073H}, 
the collisional ionization \citep{janev1987elementary, 1997NewA....2..181A}, 
the dielectronic recombination \citep{1973A&A....25..137A}, 
and the photoionization \citep{1992ApJS...78..341C}.

In order to determine the temperature of gas, we consider following radiative 
cooling and heating processes: 
the photoionization heating,
the collisional ionization cooling,
the dielectronic recombination cooling,
the collisional excitation cooling \citep{1992ApJS...78..341C}; 
the bremsstrahlung cooling \citep{1994MNRAS.268..109H}; 
and the inverse Compton cooling \citep{1986ApJ...301..522I} by 
assuming the cosmic microwave background radiation at $z = 0$.
Collisions between gas and dust grains and heating due to photoejection from grains \citep{Yorke1996} 
are not included in our simulations. 
We integrate the energy equation of gas implicitly as described in \citet{OYU12}. 


\subsection{Dust}

We include absorption and thermal emission of photons by dust grains in our simulations. 
To convert mass density to number density, we assume a graphite grain whose size and density are 
0.1\,\micron\,and 1.0\,g\,cm$^3$, respectively, as a typical dust particle \citep{draine2010physics}. 
Dust temperature is determined by the radiative equilibrium, and thus  
the dust temperature is independent from gas temperature. 
We assume that the dust sublimation temperature is $1500$~K; however, 
dust never be heated to this temperature in our simulations. 
We do not include photon scattering by dust grains for simplicity. 
Neglecting this process may overestimate the radiation pressure on dust grains as we will discuss later. 

\subsection{Time stepping}

Since we have to solve the static radiative transfer equation, 
the chemical reaction, and energy equations for gas and dust 
simultaneously. 
We thus iteratively solve these equations \citep{OYU12, tanaka15}
until the relative difference in the electron number density, 
$n_\mathrm{e}$, and in the dust temperature, $T_\mathrm{d}$  
in all cells become smaller than 0.5 \%. 

For this implicit time integration, we employ a timestep that is defined by the 
time-scale of the chemical reactions: 
%
\begin{equation}
  \Delta t_{\textrm{chem}, k} = 0.1 \left|\frac{n_\mathrm{e}}{\dot{n}_\mathrm{e}}\right|_k 
  + 1\times10^{-3} \left|\frac{n_\mathrm{H}}{\dot{n}_\mathrm{H}}\right|_k 
  \label{eq:chemtime-def}
\end{equation}
where the subscript, $k$, denotes the cell number. The second term in the right-hand side prevents
the timestep from becoming too short when the medium is almost neutral. 
We follow the evolution of the system with the minimum of the individual chemical timestep,  
$\Delta t_\mathrm{chem} = \min(\Delta t_{\mathrm{chem}, k})$,  
if this timestep is shorter than a timestep defined by the Courant-Friedrichs-Lewy (CFL) condition.  

\subsection{Hydrodynamics}

Hydrodynamics is solved by using a scheme called AUSM$+$ \citep{1996JCoPh.129..364L} 
in the second order accuracy in space and time. 
In order to prevent cell density from becoming zero or a negative value, 
we set the minimum number density, $n_\mathrm{H} \simeq 10^{-10}$~cm$^{-3}$. 
We have confirmed that our results are not sensitive to the choice of the threshold 
density as long as the threshold density is sufficiently low. 
Throughout this paper, we assume that dust and gas are dynamically tightly coupled. 
We have performed test simulations described in \cite{Bisbas2015} and confirmed that
our code reproduces their results. 

\subsection{Relative importance of radiation pressure and thermal pressure}

  \citet{krumholz09} introduce a parameter, $\zeta$, for quantifying the relative importance 
  of radiation pressure and thermal pressure. 
  The parameter is defined as 
%
\begin{equation}
  \zeta = \frac{r_\mathrm{ch}}{r_\mathrm{St}}, 
  \label{Kl}
\end{equation}
where $r_\mathrm{ch}$ is the radius at which the thermal pressure and the radiation pressure 
forces on an expanding shell are equal and $r_\mathrm{St}$ is the Str\"omgren radius 
calculated for the initial density distribution. 
For $\zeta > 1$, the expansion becomes radiation pressure dominated. 
In general, the value of $\zeta$ increases with luminosity of a radiation source 
and a dust-to-gas mass ratio. 
We estimate $\zeta$ for each simulation to compare 
our numerical results with the analytic predictions.
To calculate $r_\mathrm{ch}$, we need to know how many times on average a photon 
is absorbed or scattered in a shell, $f_\mathrm{trap}$ \citep{krumholz09}.
We estimate this value by an iterative procedure and obtain $f_\mathrm{trap} = 1$ 
for all our simulations.

\section{Simulation setup}
\label{sec:Results} 

\begin{table*}
  \centering
  \caption{
    Initial conditions and numerical setup for simulations. The radius and the total mass of 
    each cloud are indicated by $r_\mathrm{cloud}$ and $M_\mathrm{total}$. 
    The number densities, $n_\mathrm{H}$, $n_\mathrm{He}$, and $n_\mathrm{d}$ indicate the initial
    number densities of hydrogen, helium, and dust in the innermost cell, respectively.  
    The initial temperature of gas and dust are represented by $T_\mathrm{g}$ and $T_\mathrm{d}$, 
    respectively. 
    The mass of a central radiation source, which determines the luminosity, is indicated by 
    $M_\mathrm{star}$. 
    The dust optical depths from the centre to $r_\mathrm{cloud}$ at $3.29\times 10^{15}$ and 
    $1.76\times 10^{13}$~Hz are, respectively, shown as $\tau_\mathrm{d, UV}$ and $\tau_\mathrm{d, IR}$.  
  }
\begin{tabular}{lccccccccccc} \hline
  Cloud & $r_\mathrm{cloud}$ & $M_{\mathrm{total}}$ & ${n}_{{\mathrm{H}}}$ & ${n}_{{\mathrm{He}}}$ & ${n}_{{\mathrm{d}}}$ & ${T_{\mathrm{g}}}$ & ${T_\mathrm{d}}$ & ${M}_{{\mathrm{star}}}$ & $\tau_{\mathrm{d,UV}}$ & $\tau_{\mathrm{d,IR}}$ & $\zeta$ \\ 
   & (pc) & ($M_{{\sun}}$) & (cm$^{-3}$) & (cm$^{-3}$) & ($10^{-9}$~cm$^{-3}$) & (K) & (K) & ($10^3~M_{{\sun}}$) & & &  \\ \hline
  Cloud 1 &17 & 10$^5$ & 796 & 67 & 0 & 1074 & 10 & 2 & 0 & 0 &  0.13\\ 
  Cloud 2 &17 & 10$^5$ & 791 & 67 & 2.9 & 1082 & 10 & 2 & 22 & 0.15 &  0.54\\ 
  Cloud 3 &17 & 10$^5$ & 761 & 64 & 19 & 1134 & 10 &  2  & 146 & 1.0 &  4.5\\ 
  Cloud X2 &17 & 10$^5$ & 791 & 67 & 2.9 & 1082 & 10 & 20 & 22 & 0.15 &  5.0\\ 
  Cloud X3 &17 & 10$^5$ & 761 & 64 & 19 & 1134 & 10 &  20  & 146 & 1.0 &  43\\ \hline
   \end{tabular}
  \label{how}
\end{table*}

\begin{figure*}
\centering 
\includegraphics[width=16.0cm]{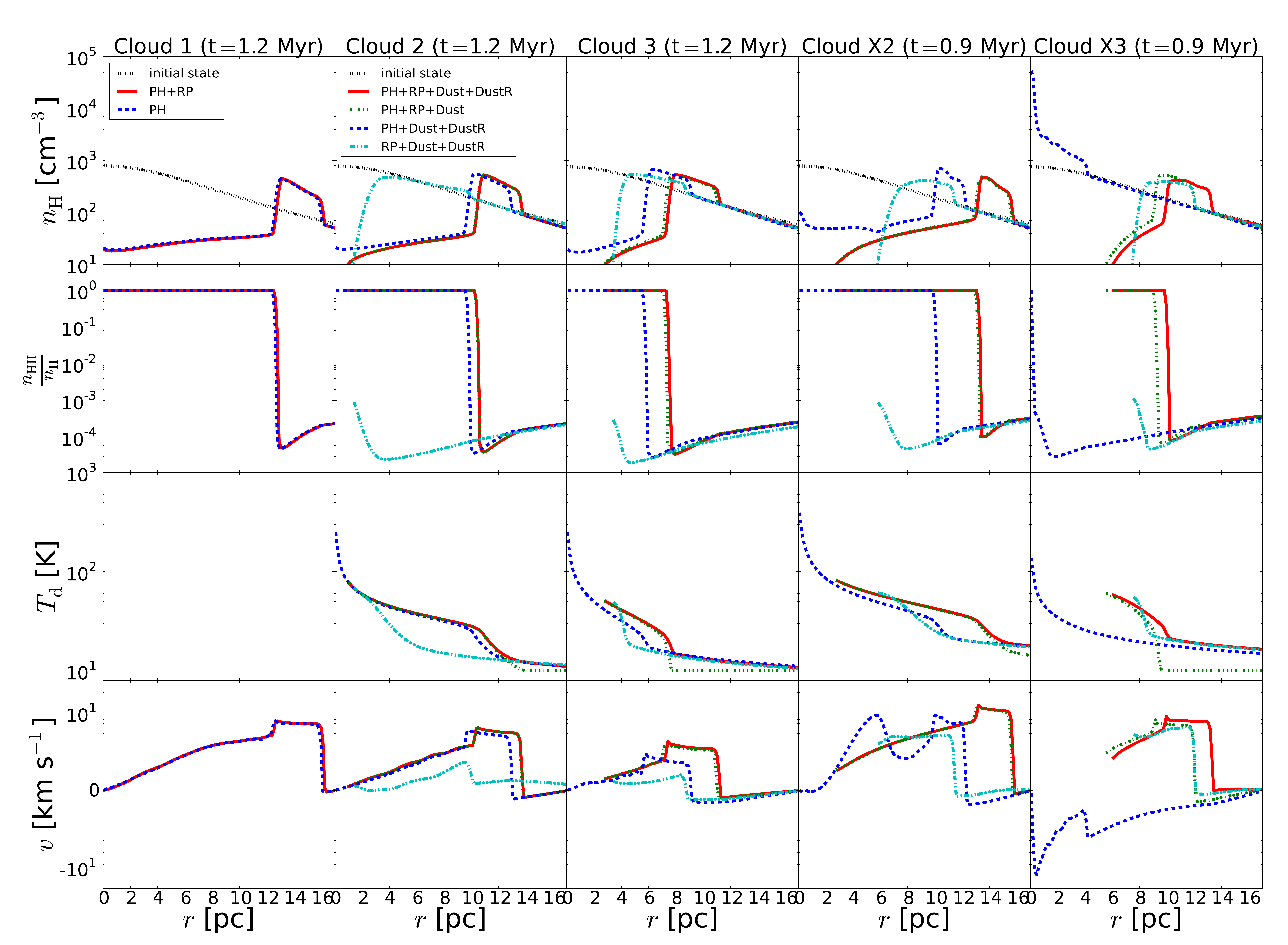} 
\caption{
  Density (top), ionization fraction (second from the top), dust temperature (second from the bottom), 
  and velocity (bottom) profiles at $t = 0.2$~Myr.  
  From left to right, we show the results for Clouds~1, 2, 3, X2, and X3. 
  The black dotted lines in the top panels indicate the initial density profiles. 
  The red solid lines represent the results of simulations that include  all radiative 
  transfer effects (`PH+RP' for Cloud~1 and `PH+RP+Dust+DustR' for Clouds~2 ,3, X2 and X3).  
  The blue dashed lines show the results of simulations in which we ignore 
  radiation pressure (`PH' for Cloud~1 and `PH+Dust+DustR' for Clouds~2, 3, X2, and X3).  
  For Clouds~2, 3, X2, and X3, we perform simulations in which we include radiation pressure 
  but we do not include absorption of re-emitted photons (`PH+RP+Dust'; green dot--dashed lines). 
  Hydrogen and helium do not interact with photons in simulations labelled as `RP+Dust+DustR', 
  and hence photoheating is disabled in these simulations (cyan double dot--dashed lines). 
} 
\label{starforming} 
\end{figure*}

To study radiation feedback in star-forming clouds,  
we model each cloud as a Bonnor--Ebert sphere of mass $10^{5}~M_{\sun}$. 
The radius of a cloud is set to obey the Larson's law \citep{larson81}. 
This gives the radius of a cloud to be 17~pc. 
As a radiation source, we place a simple stellar population (SSP) at the centre 
of the sphere. The SSP has the solar metallicity. 
We vary the mass of a radiation source for testing the role of source luminosity.
We compute its luminosity and spectral-energy distribution as functions of time by 
using a population synthesis model, {\small \textsc{P\'{E}GASE.2}} \citep{1997A&A...326..950F, 1999astro.ph.12179F}, 
assuming the Salpeter initial mass function \citep{salpeter}. 
We use linearly spaced 128 meshes in radial direction, 128 meshes in angular direction, and 256 meshes 
in frequency direction to solve radiation hydrodynamics. 
We use reflective boundary conditions at the inner boundary and semi-permeable boundary condition at the outer boundary.

Materials at radius, $r$, feel the radial gravitational acceleration, 
\begin{equation}
a_\mathrm{g} = 
-G \cfrac{M(<r)}{r^2} - G \cfrac{M_{\mathrm{star}}}{\mathrm{max}\left(r^3,  r^3_{\mathrm{soft}}\right)}r 
\end{equation}
where $M(<r)$ represents the total mass of gas inside $r$  
and $M_{\mathrm{star}}$ is the mass of the central radiation source. 
The gravitational force due to the radiation source is softened for numerical 
stability by introducing the softening length $r_\mathrm{soft}$, which is set 
to 0.5~pc.

To study the importance of dust in radiation feedback, we use 
five initial conditions, 
Clouds~1, 2, 3, X2, and X3 in which 
we vary the dust mass fraction and mass of the radiation source as follows.
Cloud~1 is a dustless cloud, while Clouds~2 and X2 have the solar 
metallicity\footnote{We employ the solar 
metallicity by \citet{asplund09}.} and we assume that half of the metals are in dust.  
Although Clouds~2 and X2 have a typical metallicity of star-forming clouds, 
its IR optical depth from the cloud centre to the edge, 
$\tau_\mathrm{IR} = 0.15$ ($\sigma_{\mathrm{IR}}=2.3\times10^{-12}$ cm$^{-2}$), 
is much lower than the value of typical star-forming clouds 
($\tau_\mathrm{IR} \sim 1$; \citealt{agertz13}) because of the low central 
concentration of a Bonnor-Ebert sphere.  
We therefore apply a higher metallicity for 
Clouds~3 and X3 so that the IR optical depths of the clouds become 
unity. 
The initial mass of the radiation source is 2~\% for Clouds~1, 2, and 3, 
while 20~\% for Clouds~X2, and X3.
The details of initial conditions are listed in Table~\ref{how}.

\section{Results}

We present density, ionization fraction, dust temperature, and velocity profiles of each cloud 
in Fig.~\ref{starforming}. 
In order to investigate the relative importance of each process, we perform simulations in 
which several physical processes are switched off. 
Simulations that include effect of increased thermal pressure due to photoheating are indicated by a label `PH'. 
When simulations do not have this label, hydrogen and helium are transparent 
for photons (photoionization and photoheating are switched off).
Simulations in which we consider radiation pressure are labelled `RP'; 
in the simulations labelled `RP', radiation pressure on hydrogen, helium, and dust is all 
included. 
Clouds~2, 3, X2, and X3 have dust, and thus simulations for these clouds have a label, `Dust'. 
In some simulations that include dust, we ignore absorption of re-emitted photons from dust. 
The label `DustR' indicates that dust can absorb re-emitted photons and hence multiple events of 
absorption and re-emission are enabled.  
Simulations that include dust and all radiative processes are named `PH+RP+Dust+DustR'.

By comparing simulations of Cloud~1, `PH' and `PH+RP', we confirm earlier results by \citet{sales14}, 
that is, the effect of radiation pressure is negligible in dustless clouds. 
Radiation pressure is also negligible in Cloud~2 as expected from the value of $\zeta = 0.54$. 
The shell expansion in this cloud is almost identical to that in Cloud~1. 
In Cloud~3 where the dust-to-gas mass ratio is increased, the effect of radiation pressure becomes 
visible. 

To isolate the effect of radiation pressure on dust, we run simulations in which  
we ignore the photoheating (and photoionization) of hydrogen and helium 
(`RP+Dust+DustR'). 
By comparing `PH+Dust+DustR' and `RP+Dust+DustR' in Cloud~3,  
we find that thermal pressure plays a more important 
role than radiation pressure in Cloud~3 in spite of the large value 
of $\zeta$. 

Adopting a higher star formation efficiency, i.e. higher source luminosity, also 
increases the relative importance of radiation pressure.
In Clouds~X2 and X3, radiation pressure is more important than in Clouds~2 and 3, 
respectively. 
In particular, thermal pressure is negligible in Cloud~X3 compared with 
radiation pressure. 
Since we increase the mass of the radiation source, thermal pressure force alone 
cannot compete the gravitational force (see `PH+Dust+DustR'); shell expansion 
is driven almost solely by radiation pressure in this case (see `RP+Dust+DustR'). 

We then investigate the impact of absorption of re-emitted photons by dust by 
comparing `PH+RP+Dust' and `PH+RP+Dust+DustR'. 
We find that this effect is negligible in almost all clouds. 
Only in Cloud~X3, radiation pressure is slightly enhanced by this process. 
Since Clouds~3 and X3 have the same IR optical depth, the importance of absorption 
of re-emitted IR photons should depend not only on the IR optical depth but also on 
the source luminosity.  

  For a given luminosity of a radiation source, a higher dust-to-gas mass ratio 
increases importance of radiation pressure. 
We, however, find that the net effect of radiation feedback (i.e. radiation pressure plus photo-heating) 
is decreased by the 
increased  dust-to-gas mass ratio; the shell radii in Clouds~1, 2, and 3 become 
smaller in the increasing order of the dust-to-gas mass ratio. 
The shell radius in Cloud~X3 is also smaller than that in Cloud~X2. 
Since the shell expansion in Cloud~X3 is dominated by radiation pressure, 
radiation feedback might become stronger than in Cloud~X2 by increasing 
the dust-to-gas mass ratio further. Doing that would enhance the radiation 
pressure via multiple events of absorption of re-emitted photons. 
The adopted dust-to-gas mass ration for Cloud~X3 is, however, already unrealistically  
high and, therefore, such a high dust-to-gas mass ratio would not be realised.  

\section{Discussion and Conclusions}
\label{sec:Discussion}

We have investigated radiation feedback in dusty clouds of radius 17~pc by one-dimensional 
radiation hydrodynamic simulations. 
In order to treat recombination radiation and re-emission from dust, 
we utilize the impact parameter method for radiation transfer.

We find that radiation pressure is negligible in a dustless cloud as pointed out by \citet{sales14}. 
Radiation pressure is almost negligible when we adopt the solar metallicity and a low star formation 
efficiency (2~\%: Cloud~2). 
This result seems to support the idea proposed by \citet{krumholz09}, that is, shell expansion is 
mainly driven by thermal pressure when the parameter, $\zeta$, is smaller than unity. 
By increasing a dust-to-gas mass ratio, the importance of radiation pressure is increased. 
Although values of $\zeta$ in Cloud~3, and X2 are significantly larger than unity, 
thermal pressure is still more dominant than radiation pressure in driving shell expansion. 
In all cases, radiation feedback creates a high density, neutral, expanding shell, which may trigger 
succeeding star formation \citep{hosokawa06}.  

We also find that effect of absorption of re-emitted photons is negligible in almost 
all clouds. 
Only in Cloud~X3, radiation pressure is slightly enhanced by this process. 
We conclude that radiation pressure cannot be significantly boosted by this 
process on cloud scale 
unless either the star formation efficiency or the dust-to-gas mass ratio  
is extremely high. 

In our simulations, radiation feedback becomes weaker for a given source 
luminosity as the dust-to-gas mass ratio increases by suppression of  
photoheating. 
This result is inconsistent with the assumption commonly made in cosmological 
simulations, that is, radiation feedback becomes stronger with the IR optical 
depth due to multiple events of absorption and re-emission of IR photons. 
\citep[e.g.][]{hopkins11, aumer13, agertz13, okamoto14}. 
Our simulations are, however, on cloud scale and the IR optical depth 
is unity at maximum. 
\citet{krumholz09} estimate $\zeta$ in star burst galaxies and 
they find, in some cases, $\zeta$ exceeds 1000. 
For such a large value of $\zeta$, radiation energy would be efficiently 
converted into radiation pressure, and radiation feedback might become 
stronger for a larger IR optical depth. 
To test this we have to perform radiation hydrodynamic simulations for 
star burst galaxies. 

Our simulations likely overestimate the impact of radiation feedback by three reasons. 
First, we model a star-forming cloud as a Bonnor--Ebert sphere. In reality, however, 
star-forming clouds are highly turbulent and characterized by self-similar fractal structure
\citep{falgarone91, elmegreen96, stutzki98}. 
Photons preferentially escape through low density medium due to the anisotropy of the thermal radiation field 
when a cloud has such complex density structure, and thus dust obtains less momentum 
compared with that in a spherically symmetric cloud \citep{Kuiper2010b,Kuiper2011,Kuiper2012,Kuiper2013}.  
In order to properly deal with this situation, we should perform three-dimensional 
radiative hydrodynamics simulations that include re-emission from dust, 
which are currently computationally too expensive (but see \citealt{Kuiper2010a}).  
Secondly, we assume that gas and dust are tightly coupled. 
Although this assumption is commonly made \citep[e.g.][]{1993ApJ...410..701N}, 
dust would leave gas behind at the shock front because 
dust grains obtain large momentum from photons and create sharp shocks in our simulations.  
If this had happened, the net impact of radiation pressure on gas 
would become weaker than in our simulations. 
Finally, we do not include photon scattering by dust grains. In reality, grains are moderately 
reflective and strongly forward scattering in UV \citep[see][]{draine03}. 
The forward scattering of UV photons could strongly decrease the radiative pressure feedback. 
We defer these issues to future studies.

\section*{Acknowledgements}

We thank the referee, Rolf Kuiper, for careful reading of our manuscript and for useful comments. 
We are grateful to Takashi Kozasa and Takashi Hosokawa for helpful discussion. 
To acknowledge the financial support of Japan Promotion of Science Grant-in-Aid for Young 
Scientists (B:24740112) and MEXT KAKENHI Grant (16H01085).




\bibliographystyle{mnras}
\bibliography{database4}

\bsp	
\label{lasPHage}
\end{document}